\documentclass[english,preprint,nofootinbib]{revtex4}
\usepackage[T1]{fontenc}
\usepackage[latin9]{inputenc}
\makeatletter

\providecommand{\LyX}{L\kern-.1667em\lower.25em\hbox{Y}\kern-.125emX\@}
\DeclareRobustCommand*{\lyxarrow}{%
\@ifstar
{\leavevmode\,$\triangleleft$\,\allowbreak}
{\leavevmode\,$\triangleright$\,\allowbreak}}

\@ifundefined{textcolor}{}
{%
 \definecolor{BLACK}{gray}{0}
 \definecolor{WHITE}{gray}{1}
 \definecolor{RED}{rgb}{1,0,0}
 \definecolor{GREEN}{rgb}{0,1,0}
 \definecolor{BLUE}{rgb}{0,0,1}
 \definecolor{CYAN}{cmyk}{1,0,0,0}
 \definecolor{MAGENTA}{cmyk}{0,1,0,0}
 \definecolor{YELLOW}{cmyk}{0,0,1,0}
}

\makeatother

\usepackage{babel}
\begin{document}
\preprint{}
\title{The origin of the KLT relations and nonlinear relations for Yang-Mills amplitudes}
\author{Luiz Antonio Barreiro}
\email{luiz.a.barreiro@unesp.br}


\thanks{}
\affiliation{S\~ao Paulo State University (UNESP), Institute of Geosciences and Exact Sciences, Rio Claro}
\author{Ricardo Medina}
\email{rmedina@unifei.edu.br}

\affiliation{Instituto de Matem\'atica e Computa\c{c}\~ao, Universidade Federal de Itajub\'a, 37500-903 Itajub\'a, Brazil}
\begin{abstract}
By means of a kinematic analysis of tree level graviton amplitudes we find, at least through six points, that the reason of their
decomposition as a sum over products of gauge boson amplitudes is on-shell gauge invariance and unitarity. As a by-product of our analysis we find nonlinear relations obeyed by Yang-Mills amplitudes. All our results are valid for arbitrary spacetime dimensions.
\end{abstract}
\maketitle

\section{Introduction}

Perturbative gravity is related to perturbative gauge theory. Since the pioneering work in \cite{Bern2} this relation is being called {\it the double copy}. Although nowadays the term {\it double copy} goes far beyond reproducing pure graviton amplitudes from gauge boson ones (see \cite{Bern5} for a recent review), at this moment the field of physical applications of the double copy is certainly gravity. In fact, there have already been results of interest that come from the double copy in gravitational wave physics (see, for example, Refs.\cite{Goldberger1, Shen1, Cheung2, Bern6}), which has become a very active field of research since the direct detection of gravitational waves in the LIGO and VIRGO experiments \cite{Ligo1, Ligo2}.

In its most simple form, at tree level, by complementing the double copy construction with color/kinematics duality \cite{Bern1}, it may be proved that it reproduces the well known Kawai-Lewellen-Tye (KLT) relations for graviton amplitudes \cite{Kawai19861, Bjerrum3, Perelstein1}. These relations not only link but state that a graviton amplitude is explicitly given as a sum of products of color ordered Yang-Mills (YM) amplitudes. In this format the diffeomorphism invariance of a graviton amplitude is automatically guaranteed by the gauge invariance of the Yang-Mills amplitudes. But the opposite is a quite a non trivial thing to prove, that is, given that a graviton amplitude is diffeomorphism invariant then it should be possible to write it as a sum of products of gauge theory amplitudes. This is something that we address in the present paper, finding the origin of the KLT relations from a pure Quantum Field Theory (QFT) perspective, although we do not need at any moment to consider a lagrangian. First of all, the original derivation of the KLT relations was done in the context of string theory \cite{Kawai19861}. Within that context its explicit generalization to the $N$-point case was later done in \cite{Bjerrum3, Perelstein1} and recently it has been found a mathematical explanation for them in intersection theory \cite{Mizera1}. Second, non string theory derivations of the KLT relations (which do not rely on Feynman diagram calculations) have been found in the recent decade. One is peculiar to $D=4$ \cite{Bjerrum4}, based on a Britto-Cachazo-Feng-Witten (BCFW) analysis \cite{BCFW}, and the other one is the Cachazo-He-Yuan (CHY) construction, valid for arbitrary spacetime dimensions \cite{Cachazo1}. Third, although these derivations succeed in reproducing the graviton amplitudes, the same as the double copy construction does \cite{Bjerrum3}, non of them explain what is the origin of the decomposition of the graviton amplitude as a sum of products of gauge theory ones, from first principles.

In this Letter our first result consists in proving that the origin of the sum over products of gauge theory amplitudes in the KLT relations is on-shell gauge invariance and unitarity.
We apply our approach to scattering processes which also consider interactions of a massless antisymmetric field and a massless scalar (the dilaton), arriving to the KLT relations involving all three fields \cite{Kawai19861} (which have the same format as the ones for gravitons).

Our second result of this paper consists in proving that, for more than four legs, there exist a set of nonlinear relations obeyed by YM color ordered amplitudes.
Nonlinear relations obeyed by YM amplitudes found in the literature \cite{Bjerrum2010268} are only valid in $D=4$, while the ones we find here are valid in arbitrary spacetime dimensions. The existence of these relations for two different given sets of BCJ basis implies that there is not a unique momentum kernel which could be used in the explicit form of the KLT relations \cite{Bjerrum3}.


Our procedure is based on a kinematic space approach first considered in \cite{Barreiro2014870} to find a kinematic derivation of the Bern-Carrasco-Johansson (BCJ) relations \cite{Bern1}, then generalized in \cite{Boels2017061602} to scattering processes involving only gravitons and only gluons, and then extended to scattering processes which consider different massless bosons, in \cite{Boels2}. This is different from other approaches which deal with gauge invariance and BCJ relations \cite{Arkani-Hamed1, Rodina1, Du1, Plefka1}.

All our results have been found through six points.

\section{Graviton amplitudes in General Relativity}


\subsection{General kinematic structure and physical requirements}

Let ${\cal M}_N$ denote the $N$ graviton on-shell amplitude in General Relativity, computed with respect to Minkowski spacetime.

${\cal M}_N$ is a Lorentz invariant kinematic expression which is multilinear in the $N$ polarization tensors $Z^{\mu \nu}_j$ and depends on the $N$ momentum vectors $k^{\mu}_j$\footnote{By demanding momentum conservation ${\cal M}_N$ may be written in terms of only $(N-1) \ k^{\mu}_j$'s.}. We demand ${\cal M}_N$ to obey the following conditions:
\begin{eqnarray}
\begin{array}{cl}
\mbox{1.}   &   k_j^2=0 \mbox{ \ \ (on-shell condition) ,} \\
\mbox{2.}   &  Z^{\mu \nu}_j=Z^{\nu \mu}_j, \ Z^{\mu \nu}_j \eta_{\mu \nu} =0 \mbox{ \ \ (symmetry and traceless condition) ,} \\
\mbox{3.}   &  Z^{\mu \nu}_j k^j_{\nu} = 0 \mbox{ \ \ (transversality or on-shell gauge condition) ,} \\
\mbox{4.}   &  \mbox{${\cal M}_N$ should remain invariant under momentum conservation, \ }  \\
\mbox{ }     &   k^{\mu}_i = - \sum_{j \neq i} k^{\mu}_j . \\
\mbox{5.}   &  \mbox{${\cal M}_N$ should be on-shell gauge invariant, \ } {\cal M}_N \big|_{Z^{\mu \nu}_j  \rightarrow  \alpha^{\mu} k^{\nu}_j + \alpha^{\nu} k^{\mu}_j} = 0 \ , \\
\mbox{}      & \mbox{where $\alpha^{\mu}$ is arbitrary, except for the condition $\alpha \cdot k_j = 0$.}
\end{array}
\label{requirements1}
\end{eqnarray}


We further demand ${\cal M}_N$ to depend on a unique coupling constant $\kappa$ and, for $N \geq 4$, to obey unitarity (in the form of factorization):
\begin{eqnarray}
\label{unitarity1}
{\cal M}_N \big|_{s_{12} \rightarrow 0} \sim \hspace{8cm} \nonumber \\
\frac{ {\cal M}^{\mu \nu}_3 (Z_1, k_1; Z_2, k_2; k)
{\cal M}_{N-1 \ \mu \nu}(-k; Z_3, k_3; \ldots; Z_N, k_N) } {s_{12}} \ ,
\end{eqnarray}
\noindent where $k = - k_1 -k_2 = k_3 + \ldots + k_N$ and
\begin{eqnarray}
\label{Mmunu1}
{\cal M}^{\mu \nu}_3 (Z_1, k_1; Z_2, k_2; k) & = & \frac{\partial}{\partial Z_{\mu \nu}} {\cal M}_3 (Z_1, k_1; Z_2, k_2; Z, k) \ ,
\end{eqnarray}
\noindent where an expression similar to (\ref{Mmunu1}) holds for ${\cal M}_{N-1 \ \mu \nu}$ in (\ref{unitarity1}).


Although not necessarilly manifest, the resulting expression of ${\cal M}_N$ should be Bose invariant.

In arbitrary spacetime dimension ${\cal M}_N$ can only be built from products of scalar terms like $\{ \mbox{Tr}(Z_i \cdot Z_j), \ \mbox{Tr}(Z_i \cdot Z_j \cdot Z_k), \ldots,  (k_m \cdot Z_n \cdot k_p), (k_m \cdot Z_n \cdot Z_p \cdot k_q), \ldots  \}$,  where ``Tr" is the trace over spacetime indices.

\subsection{The three point amplitude}

A simple example of the construction of a graviton amplitude, following the prescriptions of the previous subsection, is the three point case:
\begin{eqnarray}
{\cal M}_3 & =& i \ \kappa\ \Big\{ \mbox{Tr}(Z_1 \cdot Z_2)(k_1 \cdot Z_3 \cdot k_1) + \mbox{Tr}(Z_2 \cdot Z_3)(k_2 \cdot Z_1 \cdot k_2) + \nonumber \\
&& \hphantom{ \kappa \Big\{ \  } \mbox{Tr}(Z_3 \cdot Z_1)(k_3 \cdot Z_2 \cdot k_3) + 2 (k_2 \cdot Z_1 \cdot Z_2 \cdot Z_3 \cdot k_1)+ \nonumber \\
&& \hphantom{ \kappa \Big\{ \  } 2 (k_3 \cdot Z_2 \cdot Z_3 \cdot Z_1 \cdot k_2)+2 (k_1 \cdot Z_3 \cdot Z_1 \cdot Z_2 \cdot k_3) \Big\} \ .
\label{M3}
\end{eqnarray}
In principle, this amplitude could have been constructed using all independent $\mbox{Tr}(Z \cdot Z) (k \cdot Z \cdot k)$, $(k \cdot Z \cdot Z \cdot Z \cdot k)$, $\mbox{Tr}(Z \cdot Z \cdot Z)$, $(k \cdot Z \cdot k)(k \cdot Z \cdot Z \cdot k)$ and $(k \cdot Z \cdot k)^3$ terms. But a dimensional analysis tells us that these last two sort of terms should go with a different coupling constant than the first three ones (they would come in a higher derivative theory of gravity), so their coefficient should be zero\footnote{The scalar dimensionful combination $k_i \cdot k_j$ cannot be used in massless 3-point scattering since it is zero on-shell.}. Then, demanding on-shell gauge invariance in the three legs, all coefficients can be determined up to a global factor, which in (\ref{M3}) we have chosen to be $i \kappa$.

For $N \geq 4$ the coefficients of the kinematic terms of ${\cal M}_N$ depend on the Mandelstam variables, $s_{i j}=(k_i + k_j)^2$, $s_{i jl}=(k_i + k_j+k_l)^2$, etc.

\subsection{Two important kinematic constraints obeyed by ${\cal M}_N$}

In the next lines we argue that the following two constraints hold:
\begin{eqnarray}
\mbox{the $(k \cdot Z \cdot k)^N$ and $(k \cdot Z \cdot k)^{N-2}(k \cdot Z \cdot Z \cdot k)$ terms are absent in ${\cal M}_N$.}
\label{forbidden1}
\end{eqnarray}
First of all, in (\ref{M3}) we see that ${\cal M}_3$ obeys them. Next, for $N \geq 4$, using iteratively the unitarity relation (\ref{unitarity1}) it can be inferred that the maximum power of momenta in the numerator of ${\cal M}_N$ is $2 N-4$ \cite{Boels2017061602}. This implies, precisely, that the two kinematic constraints in (\ref{forbidden1}) should be obeyed.



\subsection{Gauge invariance and unitarity as the origin of the decomposition of ${\cal M}_N$}

\label{Gauge invariance}


The first step towards finding the expression of ${\cal M}_N$, along our approach, is to write it as a linear combination of all possible independent kinematic terms respecting (except for gauge invariance) the requirements in (\ref{requirements1}) and  the two constraints that come from unitarity, in (\ref{forbidden1}). This leads to an expression which contains the number of independent coefficients shown in the second column of Table 1.

Next, after demanding on-shell gauge invariance in all external legs, many relations arise for the previous initial coefficients, enormously reducing the number of them which are still independent, according to the third column of Table 1\footnote{For $N=6$ we have arrived at those results only working numerically.}. We call ${\cal W}_N$ the resulting vector space of kinematic gravitational expressions, to which $M_N$ belongs after demanding the afored mentioned requirements.

\begin{table}[]
\centerline{
\begin{tabular}{c|c|c}
        &  Number of independent            &   Number of independent \\
$N$  &  coefficients {\it before}           &   coefficients {\it after}  \\
        &  demanding gauge invariance   &  demanding gauge invariance\\
\hline
3 & 7 & 1 \\
4 & 336 & 1 \\
5 & 27.922 & 3 \\
6 & 5.577.852 & 21
\label{table1}
\end{tabular}
}
\caption{Number of independent coefficients in the $N$-point graviton amplitude.}
\end{table}

An extremely important result that is hidden behind the data in Table 1 is that, after demanding gauge invariance,
\begin{eqnarray}
\begin{array}{l}
\mbox{terms containing $\mbox{Tr}( \mbox{product of an odd number of} \  Z's )$}  \\
\mbox{are absent in ${\cal M}_N$.}
\end{array}
\label{forbidden2}
\end{eqnarray}

A simple example of this can be seen in eq.(\ref{M3}), in which the $\mbox{Tr}(Z \cdot Z \cdot Z)$ terms are not present in the 3-point amplitude.



In order to understand the relevance of the statement in (\ref{forbidden2}) let us be more concrete and consider the factorization procedure for a given a kinematic term of ${\cal M}_4$. Let it be, $T_1= \mbox{Tr}(Z_1 \cdot Z_2)(k_1 \cdot Z_3 \cdot Z_4 \cdot k_2)$. We define the decomposition rule of each graviton polarization tensor as
\begin{eqnarray}
Z^{\mu \nu}_j & \leftrightarrow &  \zeta^{\mu}_j \otimes_{\rm S} \bar{\zeta}^{\nu}_j \ ,
\label{change}
\end{eqnarray}
where, in order to keep a traceless polarization tensor, the polarization vectors $\zeta_j$ and $\bar{\zeta}_j$ must obey $\zeta_j \cdot \bar{\zeta}_j = 0$ and, also, in order to satisfy the transversality (third) condition in (\ref{requirements1}) they should obey $\zeta_j \cdot k_j = \bar{\zeta}_j \cdot k_j = 0$, for each light-like momenta $k_j$. Notice that the tensor product $\otimes_{\rm S}$ introduced in (\ref{change}) is symmetric in the spacetime indices.
Following the rule (\ref{change}) we have that $\mbox{Tr}(Z_1 \cdot Z_2)$ may be written as $(\zeta_1 \cdot \zeta_2) \otimes_{\rm S} (\bar{\zeta}_1 \cdot \bar{\zeta}_2)$ and $(k_1 \cdot Z_3 \cdot Z_4 \cdot k_2)$ may be written as $(\zeta_3 \cdot k_1)(\zeta_4 \cdot k_2) \otimes_{\rm S} (\bar{\zeta}_3 \cdot \bar{\zeta}_4)$\footnote{$(k_1 \cdot Z_3 \cdot Z_4 \cdot k_2)$ could have equivalently been written as $(\zeta_3 \cdot \zeta_4) \otimes_{\rm S} (\bar{\zeta}_3 \cdot k_1)(\bar{\zeta}_4 \cdot k_2)$.}. Therefore, $T_1$ can be written as a tensor product of gauge boson Lorentz invariants, as $(\zeta_1 \cdot \zeta_2) (\zeta_3 \cdot k_1)(\zeta_4 \cdot k_2) \otimes_{\rm S} (\bar{\zeta}_1 \cdot \bar{\zeta}_2) (\bar{\zeta}_3 \cdot \bar{\zeta}_4)$.

The factorization procedure mentioned above can be done with all the kinematic terms of ${\cal M}_4$, except for the $\mbox{Tr}(Z \cdot Z \cdot Z)(k \cdot Z \cdot k)$ ones. In that case we have, for example, that $T_2=\mbox{Tr}(Z_1 \cdot Z_2 \cdot Z_3)(k_1 \cdot Z_4 \cdot k_2)$ can at best be written as $(\zeta_1 \cdot \zeta_2)\zeta_3^{\mu} (\zeta_4 \cdot k_1) \otimes_{\rm S} \bar{\zeta}_{1 \mu} (\bar{\zeta}_2 \cdot \bar{\zeta}_3)  (\bar{\zeta}_4 \cdot k_2)$, that is, with left and right factors which are no longer Lorentz invariant (which is the spirit of the factorization in the KLT relations).

It is not difficult to see that in the general case of ${\cal M}_N$ the only kinematic terms which cannot be factorized in terms of Lorentz invariants are the ones which contain the trace of a product of an odd number of polarization tensors (precisely the ones mentioned in (\ref{forbidden2})). Only for them it will happen that the left and right factors will have an odd number of $\zeta_j$'s and $\bar{\zeta}_j$'s and it is not possible to contract their indices in such a way that these factors are Lorentz invariants, as it happended in the previous paragraph with the $T_2$ term.

So, the important consequence of the result in (\ref{forbidden2}) is that, after demanding on-shell gauge invariance, ${\cal M}_N$ can be written as a sum of factorizable terms:
\begin{eqnarray}
{\cal M}_N & = & \sum_{r=1}^{d_N} b_r \ \sum_s c^{(r)}_s \ U^{(r)}_s(\zeta,k) \otimes_{\rm S} V^{(r)}_s (\bar{\zeta},k) \ ,
\label{MNfactorized}
\end{eqnarray}
where, due to the constraints in (\ref{forbidden1}),  the $U^{(r)}_s(\zeta,k)$'s and the $V^{(r)}_s (\bar{\zeta},k)$'s are gauge boson Lorentz invariant kinematic expressions which do not contain $(\zeta \cdot k)^N$ and $(\bar{\zeta} \cdot k)^N$ terms, respectively\footnote{At this moment the $U^{(r)}_s(\zeta,k)$'s and $V^{(r)}_s (\bar{\zeta},k)$'s need not necessarily be Yang-Mills amplitudes. See the beginning of section \ref{The KLT relations} for their appearence in the expression of ${\cal M}_N$.}. In (\ref{MNfactorized}) $d_N$ corresponds to the number in the third column of Table 1 and it is given by
\begin{eqnarray}
d_N & = & \frac{1}{2} (N-3)! \ \big[ (N-3)! + 1 \big] \ .
\label{dN1}
\end{eqnarray}
In the beginning of the next section we explain why is $d_N$ given, not by chance, by the expression above.

Also in (\ref{MNfactorized}) the $d_N$ sums in $s$ are linearized diffeomorphism invariant expressions (where the $c^{(r)}_s$'s are specific known coefficients) and the $b_r$'s are free parameters.


\section{The KLT relations}

\label{The KLT relations}

Let us consider the BCJ basis $\{ A_N(1, \rho^{(i)}_N, N-1, N)  \}$, where $\rho^{(i)}_N$ denotes the $(N-3)!$ permutations of indices $\{2, 3, \ldots, N-2 \}$ \cite{Bern1}. Based on the fact that the BCJ set is a basis of the gauge invariant gluon kinematic expressions which do not contain $(\zeta \cdot k)^N$ terms \cite{Barreiro2014870}, in \cite{Barreiro2018} we prove that the set $\{ A_i \otimes_{\rm S} \bar{A}_j \}$ is a basis for the vector space ${\cal W}_N$, introduced in subsection \ref{Gauge invariance}, where $i \leq j = 1, 2, \ldots, (N-3)!$ and where $A_m$ denotes $A_N(1, \rho^{(m)}_N, N-1, N)$. Notice that this set is $d_N$-dimensional. So in (\ref{MNfactorized}) a change of basis of ${\cal W}_N$ can be done, allowing to write ${\cal M}_N$ as
\begin{eqnarray}
{\cal M}_N  & = &  \sum_{i \leq j = 1}^{(N-3)!} \alpha^{(N)}_{i,j} \  A_N(1, \rho^{(i)}_N, N-1, N)  \ \otimes_{\rm S} \bar{A}_N(1, \rho^{(j)}_N, N-1, N)   \ .
\label{MN2}
\end{eqnarray}

\subsection{Finding the KLT relations by demanding unitarity in eq.(\ref{MN2})}

The $N=3$ case is the most simple one and its proof does not require the use of unitarity.
For this case ${\cal M}_N$ is known (see (\ref{M3})) and in (\ref{dN1}) we have $d_3=1$, so there is only one term in the right hand-side of (\ref{MN2}). By using the well known expression of $A_3(1,2,3)$, and expanding the right hand-side of (\ref{MN2}), it is quite direct to check that the $\alpha$ coefficient is given by $i \kappa / 2$\footnote{In (\ref{M3KLT}) we have used that $\bar{A}_3(1,2,3) = - \bar{A}_3(2,1,3)$.}:
\begin{eqnarray}
\label{M3KLT}
{\cal M}_3 & = &   i \ \frac{\kappa}{2} \ A_3(1,2,3) \otimes_{\rm S} \bar{A}_3(2,1,3) \ .
\end{eqnarray}

Next, we consider $N \geq 4$. For this case we use the result in eq.(\ref{MN2}) and determine the $\alpha^{(N)}_{\sigma , \tau}$ coefficients by requiring appropriately unitarity. The procedure consists in dealing with the $\mbox{Tr}(Z_1 \cdot Z_2)(k \cdot Z \cdot k)^{N-2}$ and the $\{ (\zeta_1 \cdot \zeta_2)(\zeta \cdot k)^{N-2}, (\bar{\zeta}_1 \cdot \bar{\zeta}_2)(\bar{\zeta} \cdot k)^{N-2} \}$ terms, of ${\cal M}_N$ and $\{ A_N, \bar{A}_N \}$, respectively. The coefficients of the first terms can be determined recursively (in $N$) by demanding the unitarity requirement in (\ref{unitarity1}) and also the analog relations for the remaining $s_{1j}$ Mandelstam variables ($j=3, 4, \ldots, N$). There are similar unitarity relations which allow to determine the coefficients of the second terms as well.
Then, substituing the $(\zeta_1 \cdot \zeta_2)(\zeta \cdot k)^{N-2}$ and the $(\bar{\zeta}_1 \cdot \bar{\zeta}_2)(\bar{\zeta} \cdot k)^{N-2}$ terms in the right hand-side of eq.(\ref{MN2}), and doing their tensor product to arrive at the $\mbox{Tr}(Z_1 \cdot Z_2)(k \cdot Z \cdot k)^{N-2}$ terms, these last ones can be compared with the ones from the left hand-side of that equation allowing to determine all the $\alpha^{(N)}_{\sigma , \tau}$'s.
The subtlety at this point is that the expression found for ${\cal M}_N$ in (\ref{MN2}) is an unconventional KLT relation. The conventional ones use a different BCJ basis for both gauge sectors of ${\cal M}_N$. Then, the only thing which is left to do is to change one of the BCJ basis of a gauge sector by another one by using the appropriate BCJ relations for the amplitudes \cite{Bern1}. It is then that we arrive at the known KLT relations. For example, for $N=4$ and $N=5$ our procedure leads to
\begin{eqnarray}
\label{M4KLT}
{\cal M}_4 & = & - \ i \ \Big(\frac{\kappa}{2}\Big)^2  s_{12} \ A_{4}(1,2,3,4) \otimes_{\rm S} \bar{A}_{4}(2,1,3,4) \ , \\
\label{M5KLT}
{\cal M}_5 & = & i \ \Big(\frac{\kappa}{2}\Big)^3  \big[ \ s_{12} s_{34} \  A_{5}(1,2,3,4,5) \otimes_{\rm S} \bar{A}_{5}(2,1,4,3,5)  + \nonumber \\
&& \hphantom{  i \ \Big(\frac{\kappa}{2}\Big)^3  \big[  \ }  s_{13} s_{24} \ A_{5}(1,3,2,4,5) \otimes_{\rm S} \bar{A}_{5}(3,1,4,2,5) \ \big] \ ,
\end{eqnarray}
in agreement with \cite{Kawai19861, Bjerrum2010067}.
For $N=6$ we also arrive to the known KLT relations, which is in agreement with those references. The subtlety is that, for these case, the expressions are too big and the analytic form of the KLT relations can only be checked numerically\footnote{Assuming (\ref{MN2}) to be valid for $N=7$ we have also been able to check numerically the KLT relation for this case \cite{Barreiro2018}.}.

 The details of all these calculations will be presented elsewhere \cite{Barreiro2018}.

\subsection{The KLT relations involving also the $b_{\mu \nu}$ and $\phi$ fields}

We define the theory that describes the interactions of the graviton ($h_{\mu \nu}$), the antisymmetric field ($b_{\mu \nu}$) and the dilaton ($\phi$) by requiring a unique coupling constant $\kappa$ and by demanding the unitarity requirement (\ref{unitarity1}) for the interaction of all sort of states.

The whole procedure considered for the self interactions of a graviton is also applicable now, with the only subtlety that the polarization tensor of a $b_{\mu \nu}$ particle obey $Z^j_{\nu \mu} = - Z^j_{\mu \nu}$ and the polarization tensor of the dilaton is given by $Z^j_{\mu \nu} = Z^j \eta_{\mu \nu}$, where $Z^j$ is a scalar. For these cases the tensor product $\otimes_{\rm S}$ introduced in (\ref{change}) has now to be substituted by another tensor product $\otimes$ which takes into account the antisymmetry of the corresponding states of the specific scattering process. Notice that the two constraints in eq.(\ref{forbidden1}) are still valid, taking into account the previous polarization tensors.
In \cite{Barreiro2018} we give the details of these calculations but here we mention that the final result is that the KLT relations,  involving all interactions of the three fields, still hold with the same format (\ref{M3KLT})-(\ref{M5KLT}) as it is well known \cite{Kawai19861}. We have verified this through six points.

\section{Nonlinear relations obeyed by YM amplitudes}

\label{The nonlinear}

Since in the usual format of the KLT relations different BCJ basis are used for the left and right sectors of ${\cal M}_N$, instead of (\ref{MN2}) let us write this amplitude, right from the beginning, with a different BCJ basis for the $\bar{A}_N$'s, as
\begin{eqnarray}
{\cal M}_N & = &  \sum_{i , j=1}^{(N-3)!} \delta^{(N)}_{i , j} \  A_N(1, \rho^{(i)}_N, N-1, N) \ \otimes_{\rm S} \nonumber \\
&& \ \ \ \ \ \ \ \  \bar{A}_N (\rho^{(j)}_N \{2, \ldots, k_N-1\}, 1, N-1, \rho^{(j)}_N \{k_N, \ldots, N-2\}, N) , \ \ \ \ \
\label{MN3}
\end{eqnarray}
where $\rho^{(j)}_N \{2, \ldots, k_N-1\}$ and $\rho^{(j)}_N \{k_N, \ldots, N-2\}$ denote the first $k_N -2$ and the last $N-1-k_N$ indices of the permutation $\rho^{(j)}_N$, respectively, and where $k_N=[N/2+1]$.

If we had done a change of BCJ basis in the $\bar{A}_N$'s of (\ref{MN2}), we would have arrived at a similar expression to (\ref{MN3}), but consisting of a sum over $d_N$ terms only. Given that the sum in (\ref{MN3}) is done over $(N-3)!^2$ terms ($(N-3)!^2>d_N$ for $N \geq 5$) this means that ${\cal M}_N$ has been now expanded in a set of amplitude bilinears which is linearly dependent. Therefore there is not a unique solution for the $\delta^{(N)}_{i , j}$ coefficients, so the kernel of (\ref{MN3}) is different from zero and this leads to linear relations obeyed by the amplitude bilinears, or stated in another way, this leads to non-linear identities obeyed by YM color ordered amplitudes. The number of independent identities which arise (which is equal to the dimension of the kernel of (\ref{MN3})) is the difference between $(N-3)!^2$ and $d_N$, which gives
\begin{eqnarray}
N^* & = \frac{1}{2} \ (N-3)! \ \big[ (N-3)! - 1 \big] \ .
\label{Nstar}
\end{eqnarray}
Notice that $N^*$ is different from $0$ only for $N \geq 5$.

For $N=5$ in (\ref{Nstar}) we have $N^*=1$ and the only nonlinear relation that arises from solving the kernel of (\ref{MN3})) is given by
\begin{eqnarray}
 A_{5}(1,2,3,4,5) \otimes_{\rm S} \big( \gamma_1^{(5)} \bar{A}_{5}(2,1,4,3,5) +  \gamma_2^{(5)}  \bar{A}_{5}(3,1,4,2,5) \big)+ && \nonumber \\
 A_{5}(1,3,2,4,5) \otimes_{\rm S} \big( \gamma_3^{(5)}  \bar{A}_{5}(3,1,4,2,5) +  \gamma_4^{(5)} \bar{A}_{5}(2,1,4,3,5) \big) & = & 0 \ , \ \ \ \
\label{AYM5nonlinear}
\end{eqnarray}
\noindent where
\begin{eqnarray}
\begin{array}{ccl}
\gamma_1^{(5)} & = & s_{12}  s_{34}\ ( s_{15}+ s_{45})   \ ,  \\
\gamma_2^{(5)} & = &- s_{15} s_{23} (s_{24}+s_{45})- s_{13} ( s_{15} s_{24}+(s_{23}+s_{24}) s_{45}) \ ,  \\
\gamma_3^{(5)} & = & - s_{13} s_{24}\ (  s_{15}+ s_{45}  )  \ ,  \\
\gamma_4^{(5)} & = &  s_{15} s_{23} (s_{34}+s_{45})+ s_{12} ( s_{15} s_{34}+(s_{23}+s_{34}) s_{45})  \ .
\end{array}
\label{betas5}
\end{eqnarray}

The expression of the $\gamma_j^{(5)}$'s in (\ref{betas5}) has been found using the $N=5$ BCJ relations to write $\{ \bar{A}_{5}(2,1,4,3,5), \bar{A}_{5}(3,1,4,2,5) \}$ in terms of $\{ \bar{A}(1,2,3,4,5), \bar{A}(1,3,$\\
$2,4,5) \}$ in (\ref{AYM5nonlinear}).

Writing our relation (\ref{AYM5nonlinear}) in terms of the same BCJ basis used in eq.(10) of Ref. \cite{Bjerrum2010268} we see that they do not match, although we have checked that this last relation is indeed valid. The reason for this mismatch is that in $D=4$ the set of YM 5-point amplitudes of a given helicity configuration has a 1-dimensional basis (instead of a 2-dimensional BCJ one, in arbitrary spacetime dimension)\cite{Nastase1} and this would allow to still rewrite eq.(10) of Ref. \cite{Bjerrum2010268} in our format (\ref{AYM5nonlinear}), valid in any spacetime dimension.

Both approaches to arrive at the non-linear relations obeyed by YM amplitudes, ours and the one in Ref. \cite{Bjerrum2010268}, have in common that when the gravitational theory is extended to one which is supersymmetric, the supersymmetry requirement implies restrictions in the bosonic sector alone which are responsible for the relations. In our case these requirements are the ones in (\ref{forbidden1}) while in the case of Ref. \cite{Bjerrum2010268} they are the forbidden combination of helicity configurations which appear in the identities.


For $N=6$ in (\ref{Nstar}) we have $N^*=15$ and we have verified these relations numerically by considering the $\mbox{Tr}(Z_1 \cdot Z_2)(k \cdot Z \cdot k)^{N-2}$ terms in both sides of (\ref{AYM5nonlinear}). The $\gamma_j^{(6)}$ coefficients of these nonlinear relations are simply too big. Working numerically we have seen that the degree of homogeneity of them is at least 20 (in contrast to degree 3, found in the $N=5$ case in (\ref{betas5})).

The fact that for two given different sets of BCJ basis for YM amplitudes there exist nonlinear relations which they should obey implies that when writing a graviton amplitude in terms of them there is not a unique momentum kernel \cite{Bjerrum3} for that relation.

\section{Final remarks}

In this Letter we have seen that unitarity and on-shell gauge invariance are so strong constraints that, when applied to a graviton and a gluon amplitude, allow to prove the KLT relations (at least through six points). The key ingredient has been to work in kinematic space, where unitarity immediately implies the constraints (\ref{forbidden1}) while on-shell gauge invariance then reduces drastically the number of kinematic invariants to a basis (see Table 1), formed by factorized kinematic terms. The KLT relations proceed, then, by considering a basis for the gauge boson amplitudes \cite{Barreiro2014870}, unitarity once again and the BCJ relations.

As a by-product of our kinematic analysis we have found nonlinear relations obeyed by YM amplitudes which are valid in arbitrary spacetime dimensions.

It would be interesting to apply the sort of kinematic study we have considered here to the case of loop amplitudes in some gravitational theory, especially a supersymmetric one, in which the constraints in (\ref{forbidden1}) still hold,
in order to see if it is possible to find a proof of the double copy construction \cite{Bern2}, for which evidences exist \cite{Bern2, Plefka2}.

\begin{acknowledgments}
This work has made use of the computing facilities available at the Laboratory of Computational Astrophysics of the Universidade Federal de Itajub\'a (LAC-UNIFEI). The LAC-UNIFEI is maintained with grants from CAPES, CNPq and FAPEMIG. We thank Rutger Boels for collaboration in an initial phase of this work and Oliver Schlotterer for calling our attention to Ref.\cite{Mizera1}. RM would like to thank N. E. J. Bjerrum-Bohr for e-mail correspondence.
\end{acknowledgments}

\end{document}